\documentclass[doublecol,a4]{epl2}
\usepackage{amsmath,amssymb}
\newcommand{\De}{\text{\textbf{\textsf{D}}}_{\epsilon}} 
\newcommand{\cqp}{c_{\epsilon}(q,p)}                    %
\newcommand{\Tqp}{T^{\ }_{qp}}           									%
\newcommand{\Tqpd}{T_{qp}^{\dagger}}     						%
\newcommand{\equa}[1]{Eq.~(\ref{#1})}       						%
\usepackage{graphicx,graphics,wrapfig,rotating}	
\usepackage{dcolumn}
\usepackage{bm,pstricks}
\usepackage{hyperref}
\usepackage{mathptmx}

\title{Discrepancies between decoherence and the Loschmidt echo} 
\author{Bernardo Casabone\inst{1}\and Ignacio Garc\'ia-Mata\inst{2} \and Diego A. Wisniacki\inst{1}}
\shortauthor{B. Casbone, I. Garc\'ia-Mata, D. A. Wisniacki }

\institute{                    
  \inst{1} Departamento de F\'iõsica, FCEyN, UBA, Pabell\'on 1 Ciudad Universitaria, C1428EGA Buenos Aires, Argentina\\
  \inst{1} Departamento de F\'isica, Lab. TANDAR, Comisi\'on Nacional de Energ\'ia 
  At\'omica, Av. del Libertador 8250, \\
  1429 Buenos Aires, Argentina \\
\\
  \inst{} Dated: August 14, 2009
}

\pacs{03.65.Yz}{Decoherence; open systems; quantum statistical methods}
\pacs{03.67.-a}{Quantum information} 
\pacs{05.45.Mt}{Quantum chaos; semiclassical methods}
\abstract{
The Loschmidt echo and the purity are two quantities that can provide invaluable information about the evolution of a quantum system. While the Loschmidt echo characterizes instability and sensitivity to perturbations, purity measures the loss of coherence produced by an environment coupled to the system. For classically chaotic systems both quantities display a number of -- supposedly universal -- regimes that can lead on to think of them as equivalent quantities. We study the decay of the Loschmidt echo and the purity for systems with finite dimensional Hilbert space and present numerical evidence of some fundamental differences between them.
}
\begin{document}
\maketitle
Some of the latest breakthroughs in theoretical and experimental quantum physics have permitted among other things to explore and manipulate new states of matter -- like Bose-Einstein condensates -- and also manipulate small numbers of atoms or ions making it possible to test some of the assertions of relatively new areas of research like quantum information.
Two of the main problems that affect the achievement of such advances are uncontrolled coupling to an environment  and irreversibility caused by sensitivity to small perturbations of the quantum evolution.

The presence of a 
coupling to an external bath introduces decoherence \cite{ZurekRMP}.
By definition decoherence washes out interference terms due to quantum superposition. 
 One way of characterizing the decrease of the interference terms caused by decoherence is 
 by measuring the purity of the 
system as a function of time. For classically chaotic systems it was 
conjectured \cite{zurek1994} and numerically shown\cite{zurek1994,Monteoliva2000,Petit2006} that for a certain range of values, 
the exponential decay of purity is independent of the coupling strength 
and is characterized by the Lyapunov exponent of the classical counterpart.  
Complementarily,  to 
characterize irreversibility and instability arising from the chaotic 
nature of systems the Loschmidt echo (LE) -- also known as fidelity -- has been used \cite{Peres1984,Jalabert2001,Jacquod2001,Gorin2006,Jacquod2009}.
The idea is to study the overlap as a function of time of two states evolving with 
slightly different evolution operators characterized by some perturbation parameter 
$\Sigma$. { To avoid the singularity of a particular initial condition the LE is computed
by averaging over many initial states. The averaging can be treated in 
analogy  to the effects of decoherence and it 
was claimed  \cite{Cucc2003} -- or expected \cite{Chau2009} -- that at least for classically 
chaotic systems, since they exhibit the same decay rates, they provide 
{\em essentially\/} the same information.
  Both the purity and
the LE are of capital importance in experiments because they are readily measurable. 
In quantum information fidelity helps determine the accuracy of quantum gates and state preparation. Purity on the other hand characterizes 
how much the system is coupled to an environment or how much two parties of a system are entangled with each other.}

In the present contribution the aforementioned quantities are explored 
for systems with { finite dimensional} Hilbert space, and which have a classically chaotic counterpart. 
We present numerical evidence that, contrary to previous results \cite{Cucc2003,Petit2006}, 
significant  differences {\em can exist} between the --supposedly universal--  behaviors of the LE and the purity. We will focus on
the asymptotic regime as a function of perturbation [and decoherence] strength, 
{\em both\/} in the small perturbation regime as well as for larger perturbations. Differences 
in the short time regime were already studied elsewhere [see e.g. \cite{Jacquod2004}].
While for small perturbations the LE shows the expected quadratic regime, 
we show that the decay rate of the purity, as a function of the coupling strength, 
depends --strongly-- on the type of environment affecting the system.
Furthermore,
in the strong perturbation regime, the LE presents 
an oscillatory behavior that can mask the Lyapunov decay \cite{wang2004,wang2005,andersen2006,Natalia2009}. 
A measurement of the fidelity decay in these regimes can thus give results that are 
far from the expected. Besides, the Lyapunov decay for the purity is only observed for special types of environment.

\begin{figure*}[htbp]
\begin{center}
\includegraphics[width=0.9999\linewidth]{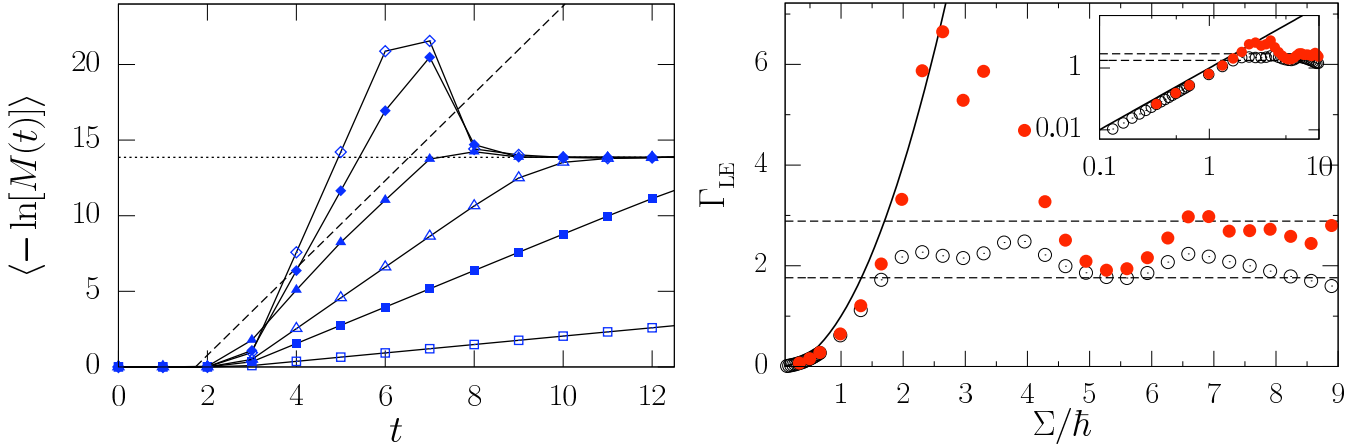}
\caption{
Left panel: The LE as a function of time for the map of Eq.~(\ref{eq:pcat}) with $a=b=4$, $N=2^{20}$ (corresponding to {\large\red $\bullet$} symbols on the right panel), and averaged over $2^{10}$ initial states uniformly distributed. The diferent values of the rescaled perturbtion are $\Sigma/\hbar=({\blue \square})\, 0.65884,\, ({\blue\blacksquare})\, 1.31768,\, ({\blue\triangle})\,1.6471,\, ({\blue\lozenge})\, 2.63536,\, ({\blue\blacklozenge})\, 2.96478,\, ({\blue\blacktriangle})\, 6.5884$. The slope of the dashed line is the Lyapunov $\lambda=\ln[9 + 4 \sqrt{5}]\approx 2.88727$. The horizontal line is at the saturation
$\ln(N)$.
Right panel: Decay rate $\Gamma_{\rm LE}$ of the LE as a function of the  rescaled strength of the perturbation $\Sigma/\hbar$.
The map is the quantum version of the perturbed cat [\equa{eq:pcat}] with ($\odot$) $a=b=2$; ({\large\red $\bullet$}) $a=b=4$. Other parameters are: $k=0.0002$, $N=2^{20}$, 
and 1024 randomly chosen initial states. 
The lines are: (dashed) $(\Sigma/\hbar)^2$. The horizontal dotted 
lines correspond to the Lyapunov exponents of the corresponding maps: (below) $\lambda=\ln [3+2\sqrt{2}]\approx 1.76275$; (above) $\ln[9 + 4 \sqrt{5}]\approx 2.88727$. The inset shows the same in log-log scale where the quadratic small-$\Sigma$ regime is best appreciated.
\label{fig1}
}
\end{center}
\end{figure*}
The systems we consider are quantum maps on the torus. { Quantum maps provide a
useful tool to understand universal properties of quantum chaotic systems -- e.g universal spectral statistics.
In addition, there exist efficient quantum algorithms for some quantum maps that can be implemented with a small number of qubits  \cite{Bertrand2001,Bertrand2004}.
This makes them ideal testbeds for current quantum computers.}
The quantized torus has associated an $N$ dimensional Hilbert space with Planck constant $\hbar=1/2\pi N$, and the position basis $\{q_i\}_0^{N-1}$ and momentum basis $\{p_i\}_0^{N-1}$ are related by the discrete Fourier transform (DFT).  We consider quantum maps $U$ whose classical counterpart are chaotic. For simplicity, we use maps whose evolution operator for one iteration can be written as follows
\begin{equation}
U=e^{i2\pi NT(p)}e^{-i2\pi NV(q)}.
\end{equation}
These kind of maps can be efficiently computed using the DFT (through the fast Fourier transform) and some have been implemented experimentally -- e.g \cite{Raizen1995}-- and have efficient quantum algorithms -- e.g. \cite{Bertrand2001,Bertrand2004}. The corresponding classical map is 
\begin{equation}
\begin{array}{lcl}
p' &=&p-\frac{dV(q)}{dq}\\
 q' &=& q-\frac{dT(p')}{dp'}
 \end{array}
\  ({\rm mod}\ 1).
 \end{equation} 
In particular for numerical calculations we use the cat map {  perturbed with a non-linear shear}
\begin{equation}
\begin{array}{lcl}
p'&=&p+a\,q +2\pi k(\cos[2\pi q]-\cos[4 \pi q])\\
q'&=&q+b\,p'
\end{array}
\  ({\rm mod}\ 1),\label{eq:pcat}
\end{equation}
with $a,b$ integers.
This map is chaotic with largest Lyapunov exponent  $\lambda\approx \ln((2+ab+\sqrt{ab(4+ab)})/2)/2$, for $k\ll 1$. { The perturbation
destroys the symmetries related to the arithmetic nature of cat maps -- which are responsible for non-generic spectral statistics. }

Let us first consider the study of the LE.
For pure states the LE is defined as
\begin{equation}
M(t)=|\langle \psi|U_{k'}^{\dag t} U_k^t|\psi\rangle|^2.
\label{eq:Mt}
\end{equation}
We define the perturbation strength as
\begin{equation}
\Sigma=k'-k.
\end{equation}  
In what follows 
$t$ is an integer that represents the number of times that the map $U$ is applied.
\equa{eq:Mt} is dubbed ``echo'' because it measures the overlap between a state 
evolved forwards up to time $t$ with $U$ and then backwards with the slightly perturbed 
operator $U_\Sigma$. It can also be seen as a measure of the separation of two, 
initially identical states, evolved forwards with two slightly different 
evolution operators. If the classical dynamics is chaotic, { there are three well identified
regimes for the LE as a function of time:
parabolic or Gaussian for very short times; 
exponential for intermediate times followed by a saturation depending on the effective Hilbert space size.
Here we  focus on the decay rate  $\Gamma_{\rm LE} $  as a function of  $\Sigma$ for the exponential decay regime.
The decay rate can be extracted from the smooth curves [see Fig.~\ref{fig1}] obtained after performing an average over 1024
uniformly distributed intial coherent states, chosen randomly.

Fig.~\ref{fig1} [left] shows the decay of the LE as a function of  discrete time $t$  for various values of $\Sigma$.  We see that after a few steps the decay is exponential. As expected for small perturbation strength (e.g. $\Sigma/\hbar=0.65884,\, 1.31768,\, 1.6471$)  the decay rate is much smaller than $\lambda$, but  contrary to predictions \cite{Jalabert2001,Jacquod2001} as we increase $\Sigma$ (e.g $\Sigma/\hbar=2.63536,\, 2.96478$) decay rates can reach values much larger than $\lambda$.  The Lyapunov decay appears for greater values of the perturbation (e.g. $\Sigma/\hbar=6.5884$). This complex behavior of the decay rate of the LE as a function of the perturbation is shown in detail in Fig.~\ref{fig1} [right] where we plot the decay rate $\Gamma_{\rm LE}$ as a function of the rescaled strength of the perturbation $\Sigma/\hbar$.  We tested results for two different versions of the map of Eq.~(\ref{eq:pcat}) with different Lyapunov exponent [($\odot$) $a=b=2$, $\lambda=\ln (3+2\sqrt{2})$; ({\large\red $\bullet$}) $a=b=4$, $\lambda=\ln(9 + 4 \sqrt{5})$].
We can see that for small perturbation strength the behavior is, as expected, 
$\Gamma_{\rm LE}\propto \Sigma^2$ -- usually called Fermi golden rule (FGR) regime. For larger perturbation strengths, 
 the decay rate is not as commonly 
predicted in the literature  [see \cite{Gorin2006,Jacquod2009} and references therein]  --with some exceptions, e.g. \cite{wang2004,wang2005,andersen2006,Natalia2009}-- perturbation independent behavior.
We find oscillations behavior near the  
value $\lambda$. These 
oscillations can be understood through the local density of states (LDOS). For finite dimensional Hilbert space the LDOS grows quadratically with the perturbation up to a point 
where it starts to oscillate. If the mean value of the oscillatory part is comparable or smaller than the classical Lyapunov exponent, then 
the oscillatory behavior is reflected in the echo. If, on the contrary, the Lyapunov exponent is much smaller  than the mean value of the 
oscillations of the LDOS, then no oscillations are appreciated in the LE \cite{Natalia2009}. The important thing to remark is that, after the 
FGR behavior, the decay of the LE is {\em not\/} perturbation independent. This can explain the difficulty to find the Lyapunov regime in echo experiments \cite{andersen2006}.

We now consider the evolution of our system in the presence of an environment. We explore the behavior of the purity for different types of environments.
Interaction between system and environment produces global state which is non-separable, i.e. entangled. 
Once we trace out the environment degrees of freedom the reduced density matrix obtained evolves non-unitarily with a consequent loss of coherence. One way to measure the effect of the decoherence produced by the environment is through the purity [see e.g. \cite{Wisniacki2009}] as a function of time 
\begin{equation}
P(t)={\rm tr}(\rho^2_t),
\end{equation}
were $\rho_t$ is the reduced density matrix of the system.
The purity is basis independent and measures the relative weight of the non-diagonal matrix elements. It can be used to measure how entangled are two systems coupled together. If  $P(t)=1$, it means that the global system can be factorized into two separate systems and there is no entanglement. On the contrary for maximally entangled states the reduced density matrix has minimum purity and the state is maximally mixed. In the case of an $N$ dimensional system  $P(t)=1/N$ for a maximally mixed state.
{ As a function of time, after an initial short transient, the purity decays exponentially. Like the LE for long times it saturates to a minimum value given by
$\hbar$. We focus on the exponential decay and the dependence of the decay rate on the coupling parameter. }

\begin{figure*}[htbp]
\begin{center}
\includegraphics[width=0.9999\linewidth]{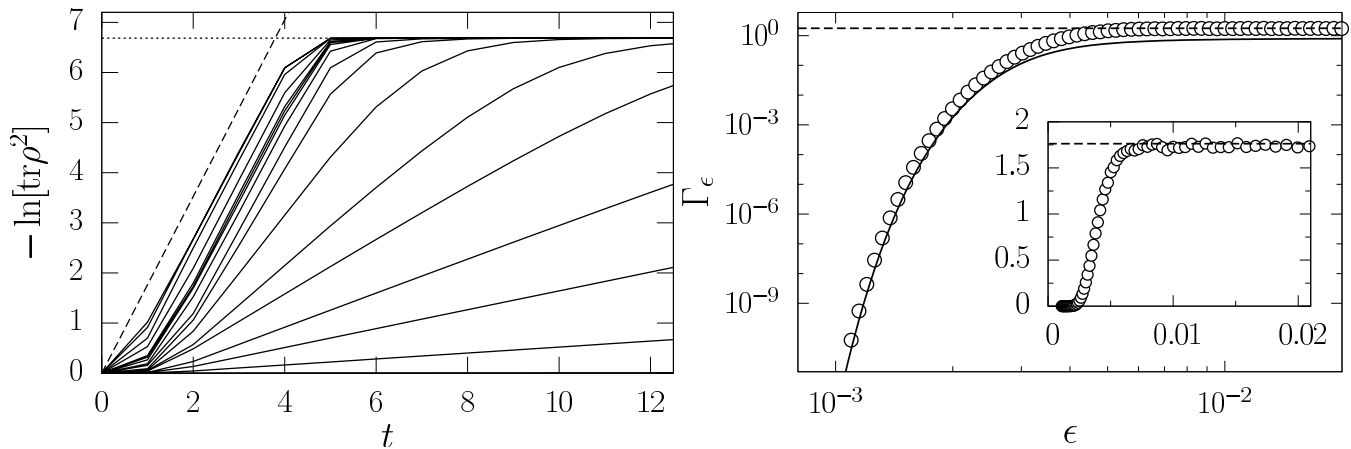}
\caption{
Left panel: Purity as a function of time for  for the map of Eq.~(\ref{eq:pcat}) with $a=b=2$, $N=2^{20}$ for various values of the decoherence parameter 
$\epsilon$. Increasing $\epsilon$ corresponds for lines from right to left (or bottom to top). The dashed stright line has a slope equal to $\lambda=\ln [(3+2\sqrt{2})]$. The dotted line shows the saturation at $\ln(N)$.
Right panel: Decay rate $\Gamma_\epsilon$ (given by the slopes of the curves in the left panel) of the purity as a function decoherence coefficient $\epsilon$ (in log-log scale) for the GDM. The map is the perturbed cat of \equa{eq:pcat} with $a=b=2$  and  $k=0.01$, $N=800$. 
The (solid) line is an approximate ($\epsilon\ll 1$) analytic calculation 
$\Gamma_\epsilon=4\frac{\exp(-2\pi^2/(\epsilon N)^2)+4\exp(-2\pi^2/(\epsilon N)^2)^2}{(1+4 \exp(-2\pi^2/(\epsilon N)^2)+\ldots)^2}$; 
The horizontal (dashed) 
line corresponds to $\lambda=\ln [(3+2\sqrt{2})]$.  The inset shows the same plot in linear scale where the $\epsilon$-independent regime with $\Gamma_\epsilon=\lambda$ is clearly observed.
All the results were obtained from the evolution of a single initial state [no average was performed].
\label{fig2}
}
\end{center}
\end{figure*}
Instead of studying the evolution of system plus environment and then tracing the environment out, we 
model directly 
the effect of the environment as a map of density matrices, or superoperator which, for Markovian environment and weak coupling,  can be written in Kraus 
operator sum form \cite{kraus}. The decoherence models we use can be expresed as a weighed sum of unitary operations, 
\begin{equation}
\rho'\stackrel{\rm def}{=}\De(\rho)=\sum_{p,q=0}^{N-1}\cqp\Tqp\rho\Tqpd ,
\label{eq:De}
\end{equation}
where $\Tqp$ are the translation operators on the torus,  $\cqp$ is a function of $q$ and $p$ and $\epsilon$
characterizes the strength.  The Kraus form implies complete positivity and the trace is preserved if $\sum_{q,p}\cqp=1$. Furthermore, as $\Tqp$ are unitary, the identity is preserved, i.e. the map $\De$ is unital.
Although position and momentum operators are not well defined in { finite dimensional} Hilbert space, translations 
can be defined as cyclic shifts \cite{schwinger}. 
In Ref.~\cite{aolita2004} it is shown that a variety of noise superoperators can be implemented in the form of Eq.~(\ref{eq:De}). The interpretation is simple: with probability $\cqp$ every possible translation in phase space is applied to $\rho$ (incoherently). 
The decoherent effect of $\De$ is evident: suppose we have a Schr\"odinger cat state that  exhibits interference fringes in the Wigner function. \equa{eq:De} written for the Wigner function of $\rho$ results
\begin{equation}
	\label{eq:wig}
W'(Q,P)=\sum_{q,p}\cqp W(Q-q,p-P).
\end{equation}
Then this incoherent sum of slightly displaced Wigner functions, washes out fast oscillating terms leaving only the classical part.

The complete map with decoherence takes place then in two steps, the unitary followed by the nonunitary part
$\rho'=\De (U\rho U^\dag)$.
This is an approximation that works exactly in some cases, e.g. a billiard that has elastic collisions on the walls and diffusion 
in the free evolution between collisions. 

To model diffusive decoherence we can define 
\begin{equation}
\label{cqp_GDM}
\cqp=\frac{1}{A}\exp\left[-\frac{q^2+p^2}{2\left(\frac{N\epsilon}{2\pi}\right)^2}\right],
\end{equation}
periodized to fit the torus boundary conditions. 
We will call this model Gaussian diffusion model (GDM). 
{ \equa{eq:wig} in the continuous limit is a convolution of the Wigner function with a kernel $\cqp$. For the GDM this corresponds to the solution
of the heat equation with diffusion constant given by $(N \epsilon/2\pi)^2$ \cite{zurek1994,Strunz1998,Carvalho2004}.
}

In Fig.~\ref{fig2} [left] we show the behavior of the purity as a function of time for map of Eq.~(\ref{eq:pcat}) with $a=b=2$, $N=800$ in the presence of GDM for different values of $\epsilon$. The exponential decay is clearly observed. Moreover, as $\epsilon$ increases (from right to left in the lines) the decay becomes independent of the value of $\epsilon$ and is given by the classical value $\lambda$.

In continuos Hilbert space and in the presence of GDM type decoherence, 
{ the decay rate of the purity  exhibits two different 
regimes as a function of the coupling parameter $\epsilon$. For small values of $\epsilon$
it is equivalent to that of the LE as a function of $\Sigma$ \cite{Cucc2003,Petit2006}, 
i.e. the decay rate $\Gamma_{\epsilon}$ 
depends quadratically for small $\epsilon$.}
Then, after a critical value 
it becomes independent of the environment and results 
$\Gamma_{\epsilon}=\lambda$ \cite{zurek1994,Petit2006}. 
 For large enough $\epsilon$ values, the behavior of $\Gamma_\epsilon$ in the case of quantum maps is the same. After a 
critical value the decay rate saturates to a constant value given by $\lambda$. 
However, as we show in Fig.~\ref{fig2} [right], for $\epsilon$ small the dependence of $\Gamma_\epsilon$ is nowhere near quadratic.
For the GDM [Fig.~\ref{fig2}, right] if $\epsilon$ is very small, of order $1/N$ then the probability of applying any translation is negligibly small. 
Thus for $\epsilon \lesssim 1/N$ there is no decoherence and the purity remains constant and equal to unity.  For larger 
decoherence strengths, the purity decays exponentially but the dependence of   $\Gamma_\epsilon$ is not quadratic. 
We remark that all the calculations done for the purity do not need any kind of averaging. 
Fig.~ \ref{fig2} was obtained using a {\em single\/} Gaussian initial state.

We can derive an
approximate analytic expression for the small $\epsilon$ regime. If we assume
$\partial_t{\rm tr}\rho^2\equiv \Delta {\rm tr}(\rho_{n+1})^2={\rm tr}(\rho_{n+1}^2)-{\rm tr}(\rho_{n}^2)=-\Gamma_\epsilon\, {\rm tr}(\rho_{n}^2$),
then from Eqs.~(\ref{eq:De}) and (\ref{cqp_GDM}), if $\epsilon\ll 1$, we have
\begin{eqnarray}
\rho'&\approx &c_\epsilon(0,0)\rho +  c_\epsilon(0,1)T_{0,1}\rho T_{0,1}^\dag+  c_\epsilon(1,0)T_{1,0}\rho T_{1,0}^\dag
\nonumber \\
& &\ +  c_\epsilon(-1,0)T_{-1,0}\rho T_{-1,0}^\dag+  c_\epsilon(0,-1)T_{0,-1}\rho T_{0,-1}^\dag.
\end{eqnarray}
We want to take the square of the trace, so the first approximation we take is ${\rm tr}(\rho T_{i,j}\rho T_{i,j}^\dag )_{i,j=0,1}\approx {\rm tr}(\rho^2)$, and we neglect higher order terms as well as higher order translations (even $T_{1(-1),1(-1)}$). 
{ We also take into account the fact $\cqp$ is symmetric around $(q=0,\, p=0)$.}
Thus we have 
${\rm tr}\rho'^2-{\rm tr}\rho^2\approx (c_\epsilon(0,0)^2+4  c_\epsilon(0,1)^2-1){\rm tr}\rho^2.$
Now, neglecting also higher order terms in the normalization [remember that $\sum_{q,p}\cqp=1$], we get
\begin{equation}
\Delta {\rm tr}\rho^2\approx 
-4\frac{\exp(-2\pi^2/(\epsilon N)^2)+4\exp(-2\pi^2/(\epsilon N)^2)^2}{(1+4 \exp(-2\pi^2/(\epsilon N)^2)+\ldots)^2}\,{\rm tr}\rho^2
\label{gammaGDM}
\end{equation}
For small $\epsilon$ we can of course neglect the terms coming from periodic boundary conditions. { This expression reproduces very well
the results obtained numerically [see Fig.~\ref{fig2}, right, solid line] in the $\epsilon \approx O(1/N)$ region.}

\begin{figure*}[htbp]
\begin{center}
\includegraphics[width=0.9999\linewidth]{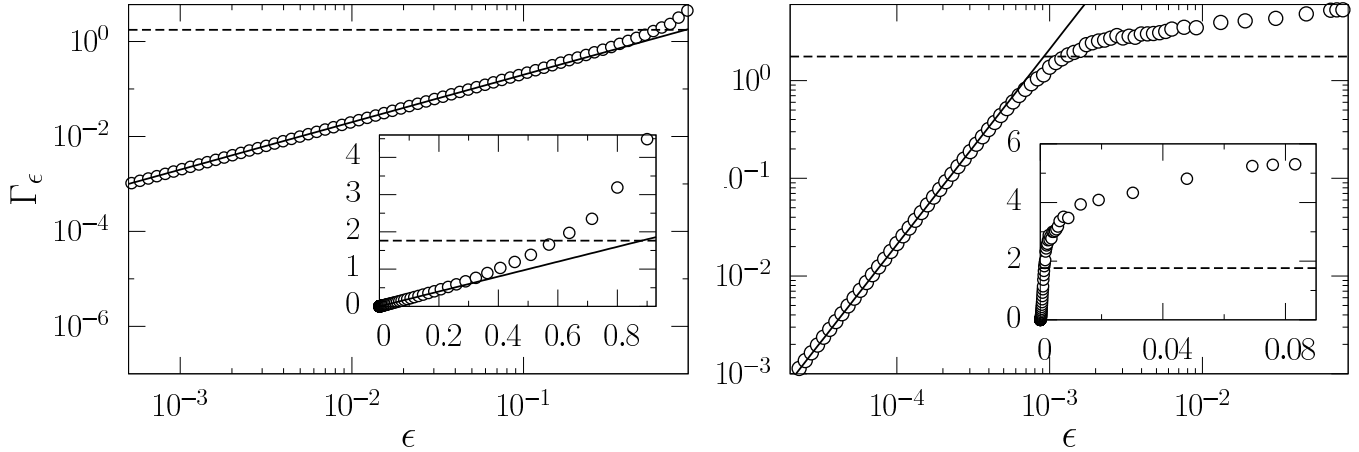}
\caption{Decay rate $\Gamma_\epsilon$ of the purity as a function decoherence coefficient $\epsilon$ (in log-log scale) for decoherence models: 
(left) DC;
(right) LDM. The map is the perturbed cat of \equa{eq:pcat} with $a=b=2$  and  $k=0.01$, $N=800$. 
The solid lines shows the fit for small $\epsilon$. 
For the DPC (left) the solid curve was obtained analyticaly and is $\Gamma_\epsilon=2\,\epsilon$. For the LDM [right] the fit is aproximately 
 $\Gamma_\epsilon\propto \epsilon^2$.
 The dashed line in both plots corresponds to  $\lambda=\ln [(3+2\sqrt{2})]$. The inset in both panels displays the same plot in linear scale where the absence of an $\epsilon$-independent Lyapunov regime is clearly observed for both DC and LDM cases.
\label{fig3}
}
\end{center}
\end{figure*}
In order to attain the quadratic dependence of $\Gamma_\epsilon$ for small coupling, observed in continuos Hilbert space, $\cqp$ should  have tails that decay slower than Gaussian, i.e. long distance correlations in phase-space. 
 We can for example take a well known decoherence channel for quantum information processing, the depolarizing channel (DC) \cite{nielsenBook}, which is also a convex sum of unitaries and can be simply written in terms of translations in phase space \cite{aolita2004}
 \begin{equation}
 \De^{\rm DC}=(1-\epsilon)\rho+\frac{\epsilon}{N^2}\sum_{q,p\ne 0}\Tqp\rho\Tqpd
 	\label{eq:dc}
 \end{equation}
 In Fig~\ref{fig3} [left] we show the decay rate $\Gamma_\epsilon$ for the DC.
 Following a similar reasoning as the one followed to obtain \equa{gammaGDM} we get, 
 for $\epsilon\ll 1$, $\Gamma_\epsilon=2\epsilon$ [see Fig~\ref{fig3}, left, solid line].
The DC is an extreme case to consider as phase space decoherence because it is highly non-local: with the same probability it implements every possible translation $\Tqp$ ($q,p\ne 0$). Therefore, there is no reason to expect a Lyapunov regime in this case. In fact for $\epsilon$ close to 1, 
the dynamics is dominated by the environment. The absence of Lyapunov regime (or any $\epsilon$ independence) is clearly appreciated in Fig~\ref{fig3} [left] .
The non-locality of DC has also devastating effects on the entangling power of the algorithms that implement chaotic maps \cite{GarciaMata2007}.

 To reproduce the FGR quadratic regime we thus need a decoherence model which is peaked at $c_\epsilon(0,0)$ and which has polynomially decaying tails. 
 We propose to take a Lorentzian
 \begin{equation}
 	\label{eq:LDM}
\cqp=\frac{1}{\pi A}\sum_{j,k=-x}^x\frac{\frac{\epsilon N}{2\pi}}{\left( \left(\frac{\epsilon N}{2\pi}\right)^2+(q-Nj)^2+(p-Nk)^2\right)}
\end{equation}
with $A$ the proper normalization for $\sum_{q,p}\cqp=1$.
We will call this case Lorentz decoherence model (LDM). The sum is done to account for the periodicity of the torus (theoretically $x\to\infty$, practically $x$ is an integer much larger than 1). 
{ \equa{eq:De} with $\cqp$ given by \equa{eq:LDM} defines a random process with Lorentzian weight. We can relate this to superdiffusion by L\'evy flights. }
Long tail decoherence was also considered in Ref.~\cite{Schomerus2007} where it was shown  
that the decoherence rates can be tuned to power law decay in cold atom experiments.
In Fig.~\ref{fig3} [right], we show $\Gamma_\epsilon$  for the  LDM. The quadratic dependence is clearly observed. As in the DC model the Lyapunov regime is not present. Larger $\epsilon$ implies longer Lorentzian tales which, when periodized sum up to non negligible non-local effects all over phase space. This is why for the LDM not only is the Lyapunov regime also not present but the decay rate of the purity continues to grow indefinitely.  { To obtain the so-called universal behavior -- quadratic-FGR growth followed by constant-Lyapunov  -- a very specific model with large tails but sufficiently localized is needed.}
A combination of both GDM and LDM, so that the former dominates at larger $\epsilon$ and the latter dominates for smaller $\epsilon$  would yield both the FGR regime and the Lyapunov regime. Decoherence combining both Gaussian and Lorentzian processes was studied e.g. in \cite{Vacchini2005}. 

To summarize, the LE and the purity for systems with { finite dimensional} Hilbert space has been analyzed. We have shown that though they can exhibit 
qualitative  similarities, they are fundamentally very different:  the small coupling regime for the purity is not quadratic but depends on the environment model. 
Moreover, while the large perturbation regime for the LE can present high amplitude oscillations around the classical Lyapunov exponent  depending on the LDOS,
 for the purity it depends decidedly on the type of environment. Only environments that act {\em locally\/} in phase space exhibit the -- independent -- Lyapunov regime. Thus, we remark that the LE and the purity provide intrinsically different information.

 The authors acknowledge financial support from CONICET (PIP-6137) , UBACyT (X237) and ANPCyT.  D.A.W.  and I. G.-M. are researchers of CONICET.  
Discussions with M. Saraceno are thankfully acknowledged.


\end{document}